**Intercomparison of micro- and nanodosimetry Monte Carlo simulations: an approach to assess the influence of different cross-sections for low-energy electrons on the dispersion of results.**

Carmen Villagrasa[1,8], Hans Rabus[2,8], Giorgio Baiocco[3,8], Yann Perrot[1,8], Alessio Parisi[4,8], Lara Struelens[4,8], Rui Qiu[5,8], Michaël Beuve[6,8], Floriane Poignant[6,8,*], Marcin Pietrzak[7,8], Heidi Nettelbeck[2,8]

[1] Institut de Radioprotection et de Sûreté Nucléaire (IRSN), Fontenay-Aux-Roses, France

[2] Physikalisch-Technische Bundesanstalt (PTB), Braunschweig and Berlin, Germany

[3] Physics Department, University of Pavia, Pavia, Italy

[4] Belgian Nuclear Research Center (SCK CEN), Mol, Belgium

[5] Department of Engineering Physics, Tsinghua University, Beijing, China

[6] Institut de Physique des 2 Infinis, Université Claude Bernard Lyon 1, Villeurbanne, France

[7] National Centre for Nuclear Research (NCBJ), Otwock-Swierk, Poland

[8] European Radiation Dosimetry Group (EURADOS) e.V, Neuherberg, Germany

[*]Present address: National Institute of Aerospace, 100 Exploration Way, Hampton VA 23666, USA

**Abstract**

An intercomparison of microdosimetric and nanodosimetric quantities simulated Monte Carlo codes is in progress with the goal of assessing the uncertainty contribution to simulated results due to the uncertainties of the electron interaction cross-sections used in the codes. In the first stage of the intercomparison, significant discrepancies were found for nanodosimetric quantities as well as for microdosimetric simulations of a radiation source placed at the surface of a spherical water scoring volume. This paper reports insight gained from further analysis, including additional results for the microdosimetry case where the observed discrepancies in the simulated distributions could be traced back to the difference between track-structure and condensed-history approaches. Furthermore, detailed investigations into the sensitivity of nanodosimetric distributions to alterations in inelastic electron scattering cross-sections are presented which were conducted in the lead up to the definition of an approach to be used in the second stage of the intercomparison to come. The suitability of simulation results for assessing the sought uncertainty contributions from cross-sections is discussed and a proposed framework is described.

1. **Introduction**

The European Radiation Dosimetry Group (EURADOS, www.eurados.org) is a non-profit association for promoting research and development as well as European cooperation in the field of ionizing radiation dosimetry. Currently comprising more than 79 European institutions and more than 700 scientists, a main strength of this network is the ability to promote intercomparisons and benchmarks on common issues in radiation dosimetry that help identify problems and avenues for their resolution. EURADOS Working Group 6 (WG6) is concerned with quality assurance for computational dosimetry ( 1) and, within WG6, the different task groups focus on the multiple aspects of dosimetry which can be addressed with computational methods, in particular and very frequently, with Monte Carlo (MC) codes.

Computational methods are indispensable tools in radiation dosimetry where the quantities of interest (e. g. organ doses in radiation protection) are often not accessible to measurement. Benchmarking of the calculations relies on comparison with experimental data obtained for simplified set-ups such as an ionization chamber in a water phantom in radiotherapy quality assurance. MC codes are often used in this context as they allow an analogous simulation of radiation transport and energy deposition with flexible choice of definition of geometrical details. The MC codes used in radiation applications use random sampling of radiation interaction processes based on interaction cross-section data that are implemented as data sets and/or model functions for the interpolation of the dependence of the cross sections on parameters such as impact energy and scattering angle. Generally, the data sets and model functions are derived from experimental data for cross sections and/or theoretical approaches. The latter play a fundamental role in the extreme cases of very high energetic (above several tens of MeV) or very low energetic particles (below 1 keV), where direct experimental determination of cross sections is not feasible.

EURADOS Task Group 6.2, which is concerned with computational micro- and nanodosimetry, began an intercomparison exercise a few years ago to evaluate and investigate the origin of the dispersion of results that can occur when previously validated Monte Carlo codes are applied to microdosimetric and nanodosimetric problems. A simple simulation setup was chosen for the investigation so that differences between results would predominantly reflect the impact of the cross-section models used in the codes. Following the philosophy of the Guide to the expression of uncertainty in measurement (GUM) that has been developed over the last two decades by the Joint Committee for Guides in Metrology (JCGM) ( 2, 3), the scatter of results could be interpreted as the contribution to the uncertainty of the simulated microdosimetric and nanodosimetric quantities (frequency distributions of imparted energy or number of ionizations) due to the different interaction cross sections used in the codes.

For the microdosimetric part of the exercise, the setup of the problems to be simulated included borderline cases testing the performance of certain features of the codes, especially those using condensed history approaches for electron transport. Results of the microdosimetric part of the intercomparison have been published before ( 4). Additional microdosimetric results obtained after this publication are presented in this current work, showing that general MC codes with condensed-history approaches are in some cases unsuitable for microdosimetric applications.

The main focus of this paper is, however, a follow-up on the preliminary nanodosimetric results presented in ( 4), where the ionization cluster size distribution (ICSD), which is the normalized frequency distribution of the number of ionizations, was to be determined in nanometer-size target spheres. The ICSDs reported by different participants showed a large dispersion among the few track-structure codes that were used by participants at that time ( 4). While different factors and features of the codes used can be the source of these differences, a plausible key explanation is the use of different cross-section values for the transport of low energy electrons including inelastic and elastic interactions. Including additional, more recently submitted results, a sensitivity analysis has, therefore, been performed on the dependence of nanodosimetric results on variations of the cross sections used in the simulations. Based on the insights from this sensitivity analysis, a revised approach for the second part of the intercomparison exercise can be proposed.

The structure of the paper is as follows: First, the rationale for the exercise is reviewed in Section 2. The new results and conclusions on the intercomparison of simulated microdosimetric spectra are presented in Section 3. Section 4 presents an update of the intercomparison of simulated ICSDs using track-structure codes, while Section 5 discusses the outcome of an analysis performed on the sensitivity of these nanodosimetric results to variations in cross-section data. Finally, Section 6 presents the proposed plan for the second part of the exercise and the envisioned methodology for assessing the contribution to the uncertainty of nanodosimetric results that arises from the cross-sections used in the codes.

## 2. Background and Rationale of the Exercise

Microdosimetry and nanodosimetry are concerned with the quantitative characterization of the stochasticity of ionizing radiation interactions at the microscopic scale of biological targets as cells and subcellular structures and how this connects to the biological effectiveness of radiation. Measurements in conventional microdosimetry and in nanodosimetry are performed with experimental devices (gas counters) that simulate micrometric or nanometric targets in condensed matter based on a density scaling principle ( 5, 6). The relevant quantities (frequency distributions of imparted energy or number of ionizations in a defined target volume) in biological matter are assessed by simulations using MC codes. Direct validation of these simulations is generally not possible, since corresponding experiments cannot be performed; the codes can be validated indirectly by benchmarking results of simulations of microdosimetric or nanodosimetric experimental setups with experimental data obtained in gas ( 7, 8, 9).

Besides general-purpose radiation transport MC codes, that generally apply the condensed-history approach to reduce computation time, track-structure codes have been developed for applications in micro- and nanodosimetry ( 10, 11). These codes use event-by-event simulation of all interactions and have been mainly developed in the frame of radiobiological modelling where significant experimental and computational research in recent decades has advanced our understanding of the various processes leading to radiation-induced effects ( 12)

In order to obtain the detailed pattern of radiation interactions on a nanometric scale, track-structure codes require interaction cross-sections for simulating electron transport over an energy range that extends down to energies of a few eV. From a purely theoretical point of view, the use of Monte Carlo simulations for electrons with kinetic energy < 1 keV and the information derived about the position of their interactions is highly questionable (13). Nevertheless, Liljequist and Nikjoo (14) demonstrated that a trajectory treatment provides a good approximation of multiple quantum scattering down to electron energies of the order of 10 eV, due to the incoherence introduced by a random-like structure of the medium and to the presence of multiple inelastic scattering.

Track structure codes include cross-sections for elastic scattering and all the inelastic interactions that lead to energy deposition in condensed matter. For low projectile energies, these cross-sections depend not only on the atomic composition of the target but also on its aggregation state (gas, liquid or solid). However, only few experimental data of these cross-sections exist to allow the validation of different theoretical approaches that describe the molecular orbitals of the target and the interactions with the electrons. For this reason, most track-structure codes used in the research field of radiobiological modelling adopt liquid water as a surrogate for all biological target materials, such as those of present in the cell nucleus, e.g., deoxyribonucleic acid (DNA).

Over the past few decades, various theoretical models have been developed for the calculation of elastic and inelastic scattering cross-sections of electrons in water at low energies. A review of these approaches is beyond the scope of this paper, but it is worth noting that benchmarks of the implemented cross-sections have been performed for all track-structure MC codes against the few available experimental cross-section data. These reference data are either experimental data for water vapor that are scaled to the density of liquid water or, in the case of inelastic scattering, data for liquid water obtained indirectly from measurements of the optical constants of liquid water (15, 16). Indeed, these measurements allow calculating the dielectric function in the optical limit of zero momentum transfer. The dielectric response is the material property that describes the inelastic scattering of low energetic charged particles in condensed matter and is a function of the energy and momentum transfer in the collision. The momentum dependence is generally introduced into the models by applying physically plausible dispersion algorithms that essentially re-distribute the imaginary part of the dielectric function (or oscillator strength) to the different ionization shells and excitation levels conserving the integral scattering form factor ("f-sum-rule") and maintaining the original fit to the experimental data (optical-model approach) ( 17).As the extension of the dielectric function to the entire momentum transfer range and the distribution of these values into the different ionization and excitation shells cannot be directly compared to any experimental data, such cross-section data sets have significant uncertainties that can introduce large dispersion between the results obtained with different track-structure codes. However, not only inelastic scattering cross-sections introduce this dispersion; elastic scattering models also use different approaches. Examples are the use of screening parameters derived from experiments to enlarge the applicability of the first Born approximation ( 18) or more theoretical models as the Dirac partial wave analysis ( 19,20). The validation of the implemented inelastic and elastic scattering cross-sections in the various MC track-structure codes can only be done by comparison to experimentally determined integral quantities at higher electron energy as stopping powers or ranges (21, 22). From the results available in the literature, it seems that in the case of electrons with kinetic energy higher than 200-300 eV, optical-data model calculations for the inelastic mean free path (IMFP) are not sensitive to the choice of the physical model used (~10%). For electrons with lower kinetic energy, however, the choice of the physical model highly influences the result.

Since the validation of different cross-section data sets implemented within track structure codes beyond what has already been achieved is not yet feasible, the objective of this work is to establish an approach to determine the contribution of these different cross-section data sets to the spread of nanodosimetric results that are currently obtained by users of track-structure codes. To do so, ICSDs calculated in nanodosimetric volumes at different distances from an electron source are deemed a suitable observable, as they reflect the stochastic description of the track. This is because take into account both inelastic cross-sections (i. e. the number of ionization interactions) and elastic scattering as elastic processes drive the shape of the track and influence the number of ionizations detected in the volume.

The first part of the exercise therefore adopted the "agnostic" Bayesian approach underlying the Guide to the expression of uncertainty in measurement (GUM) ( 2) . The GUM has been developed over the last two decades by the Joint Committee for Guides in Metrology (JCGM), and recently Supplement 6 on modelling has been published ( 3) . Without prior knowledge on factors that might make one code more reliable than the other, all simulation results have to be treated as independent "measurements" with an unknown location of the "measured" value within the probability distribution of possible values of the "measurand". In absence of correlations among the "measurements", the best estimate of the "true" value of the "measurand" is the mean of the reported values and its uncertainty can be inferred from the standard deviation of the ensemble of results. Correlations may be introduced by use of the same code by different participants, as it has been seen in ( 4).

For lack of possibility to assess the accuracy of the electrons cross-section models (total and mostly differential) by comparison with experimental data, the philosophy of the GUM will be also applied in the second part of the exercise to the cross-section data in the codes. The goal is to quantify the contribution of these different cross-section data sets to the dispersion of the nanodosimetric results obtained using track-structure codes. For this purpose, ICSDs calculated in nanodosimetric volumes at different distances from an electron source are deemed a suitable observable as they reflect the stochastic description of the track taking into account both inelastic cross-sections (i. e. the number of ionization interactions) and also elastic scattering because elastic processes drive the shape of the track and condition the number of ionizations detected in the volume. It should be noted that this paper does not present the final results on the uncertainty associated with the cross-sections of the track-structure codes in the ICSD results but rather the analysis of the intercomparison results and the sensitivity analysis that have led to the construction of the method that will be used.

### 3. First intercomparison: Microdosimetric results

In the first part of the intercomparison exercise launched in 2017, and a preliminary analysis of the results was published in (4). Participants were asked to calculate microdosimetric spectra (*zf*(*z*) vs *z*) as well as nanodosimetric ICSDs using simple geometries and a description of a low energy electron source with an electron energy distribution related to the average radiation spectrum reported by Howell et al. (23) for the decay of $^{125}$I. Aforementioned microdosimetric and nanodosimetric quantities were to be determined per single decay of the given source.

The spectrum of emitted electrons of this artificial source had only discrete energies equal to the average values given by Howell et al. (23) for the different transitions and groups of transitions. Furthermore, only the electron energies and average number of electrons per decay reported by Howell et al. (23) were used, whereas the physics of the individual transitions was ignored to simplify the simulation task. Thus, the simulations related to an artificial electron emitter that undergoes a decay in the following way: first, an electron is emitted with probability 0.936 where the energies and probabilities for their occurrence are given in Table 1. This electron corresponds to an internal-conversion electron in the decay of a real $^{125}$I nucleus (see last column in Table 1). Then, the decay is completed with the emission of additional 25 electrons with their energy random sampled from the discrete distribution given in

Table *2*. The energy distribution of these electrons corresponds to that given in ( 23) for the Auger and Coster-Kronig (CK) electrons emitted in the decay of a $^{125}$I nucleus. However, it should be noted that, in a real decay of a $^{125}$I nucleus, the Auger and CK cascades depend on the vacancy created by the IC process, whereas this is not the case for the artificial source used in the exercise.

**Table 1:** Energies and their probabilities for electrons emitted in the first step of the decay of the artificial electron source used in the intercomparison. The last column gives the corresponding transition in the decay of a $^{125}$I nucleus producing an internal conversion (IC) electron.

| Energy (eV) | Probability | Nominal transition |
|---|---|---|
| $3.65\times10^3$ | 0.797 | IC 1 K |
| $30.6\times10^3$ | 0.110 | IC 1 L |
| $34.7\times10^3$ | $2.84\times10^{-2}$ | IC 1 M,N |
| **Total yield of IC electrons per decay: 0.94** | | |

**Table 2:** Energies and their probabilities used for all electrons emitted in the second step of the decay of the artificial electron source used in the intercomparison. The third column gives the average yield and the last column the corresponding Auger or Coster-Kronig (CK) transitions in the decay of a $^{125}$I nucleus that produce electrons of this energy ( 23).

| Energy (eV) | Probability | Average yield / decay | Nominal transition |
|---|---|---|---|
| 6 | 0.147 | 3.66 | CK OOX |
| 29.9 | 0.141 | 3.51 | CK NNX |
| 32.4 | 0.438 | 10.9 | Auger NXY |
| 127 | $5.79 \times 10^{-2}$ | 1.44 | CK MMX |
| 219 | $1.06 \times 10^{-2}$ | $2.64 \times 10^{-1}$ | CK LLX |
| 461 | 0.132 | 3.28 | Auger MXY |
| $3.05 \times 10^3$ | $5.03 \times 10^{-2}$ | 1.25 | Auger LMM |
| $3.67 \times 10^3$ | $1.37 \times 10^{-2}$ | 0.34 | Auger LMX |
| $4.34 \times 10^3$ | $8.48 \times 10^{-4}$ | $2.11 \times 10^{-2}$ | Auger LXY |
| $22.4 \times 10^3$ | $5.55 \times 10^{-3}$ | $1.38 \times 10^{-1}$ | Auger KLL |
| $26.4 \times 10^3$ | $2.37 \times 10^{-3}$ | $5.90 \times 10^{-2}$ | Auger KLX |
| $30.2 \times 10^3$ | $2.61 \times 10^{-4}$ | $6.50 \times 10^{-3}$ | Auger KXY |
| **Total yield of Auger and CK electrons per decay: 24.9** | | | |

Participants had to implement this source in their simulations using the methodology of their choice. To assure that the simulations were conducted with the same electron energy spectra, participants were requested to also determine and report the spectra of imparted energy per decay when the simulation was run with electron transport switched off so that the electron energy was deposited at the location of the emitter. All the participants for which the results were reported in [ 4] and are presented in the following section for the surface source configuration reported spectra that agreed with the reference solution (see Supplementary Fig. S1) with an average energy released per decay corresponding to the one expected from the source definition (~19.5 keV). It was thus assumed that the participants correctly implemented the source.

For the microdosimetric calculation, three alternative source descriptions were used as presented in Figure 1 : 1) source in the center of a liquid water sphere of 5 µm radius (point source configuration); 2) source uniformly distributed in the volume of that sphere (volume source configuration) and 3) source uniformly distributed on the sphere's surface (surface source configuration). In all three cases, all points of the source have the same probability of isotropic emission of electrons with an energy spectrum as given in Table 1 and

Table *2*. The 5 µm radius sphere was always taken as the target for the calculation of the microdosimetric spectra and was placed within a liquid water world volume.

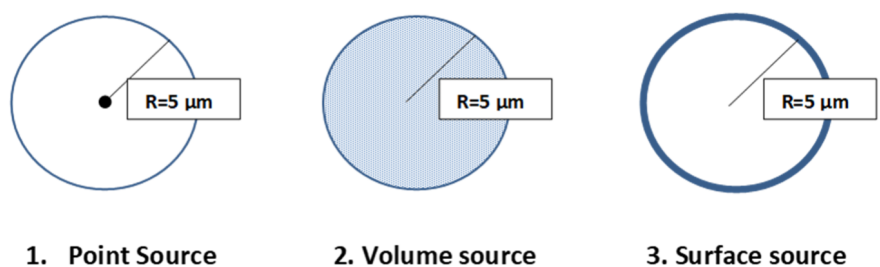

1. Point Source    2. Volume source    3. Surface source

*Figure 1: Simulation set-up used in the microdosimetric spectra benchmark. Three different configurations were proposed for the electron source position (isotropic emission). The target volume is the 5 μm radius liquid water sphere shown in the three cases. The energy spectrum source corresponds to a mean decay of $^{125}I$ as given in ( 23). For details see text.*

The rationale for these choices of source distribution was to test the performance of the codes for geometrical setups that constitute extreme cases (surface source and point source) and a situation close to a realistic irradiation scenario (volume source). For the point source, a large proportion of the energy emitted by the source is expected to be imparted in the sphere so that similar results are expected for the different codes from the principle of energy conservation. The case of the surface source, on the other hand, is sensitive to the way geometrical boundary crossing is treated in the codes and to potential deviations from isotropic emission.

At the time of publication of the first results, a total of 9 different Monte Carlo codes or cross-section data sets implemented within the same code (referred to as different "options") had been used by the participants to calculate the microdosimetric spectra. Different options in Geant4-DNA (24, 25, 26, 27, 28) and NASIC (29, 30) allow simulation of electron interactions to be done in an event-by-event basis, as is the case with track-structure codes. The rest of the codes used in the calculations published in (4) made use of a condensed history approach: Geant4-Livermore (31), PHITS (32), FLUKA (33), MCNP6 (34) and PENELOPE (35).

**Table 3** Monte Carlo codes used for the microdosimetric calculations. The energy cut-off for electron transport for each code is also indicated.

| MC code | Electron transport energy cut-off ($E_{cut}$) |
|---|---|
| **Geant4-DNA Opt0/default (V10.3)** | 7.4 eV and 14 eV |
| **Geant4-DNA Opt2 (V10.3)**[a] | 7.4 eV |
| **Geant4-DNA Opt 7 (V10.3)** | 10 eV |
| **PENELOPE** | 50 eV |
| **MNCP6.1** | 20 and 14 eV |
| **FLUKA** | 1 keV |
| **NASIC** [b] | 7.4 eV |
| **PHITS** | 1 keV |
| **PHITS etsmode** | 1 eV |
| **PARTRAC** | 10 eV |
| **Geant4-Livermore** | 1 keV (not $E_{cut}$ but production threshold) |

a) fast version of Opt0 cross-sections. b) based on Geant4-DNA Opt0 cross-sections ( 30).

Analysis of the participants' preliminary results in this microdosimetric benchmark showed a good agreement between the different MC codes for the point source configuration as well as for the volume source configuration. The agreement was assessed by means of the relative standard deviation among spectra and was found to be 2.3% and 2.8% for the point-source and volume-source configuration, respectively. The slightly higher value in the latter case was explained by the greater influence of the energy cut-off for electron transport in the codes for the volume source compared to the point source. Hence, for the considered decay energies, most of the electron ranges in water were smaller or comparable to the target sphere radius. In the case of the surface source configuration, however, large variations among the participants' results were observed. Indeed, two different groups seemed to appear from Pearson's correlation analysis (12). One group of results was obtained with PENELOPE, FLUKA and MCNP (with two energy cut-offs: 20 eV and 14 eV), while the other group included results from using three different options in Geant4-DNA, where two participants used Geant4-DNA default option with different energy cut values ($E_{cut}$ =7.4 eV and 14 eV). The remaining two results were quite different from all the others.

After this preliminary analysis and follow-up correspondence with the contributors, some results were reevaluated and new contributions from participants using different track-structure codes such as PARTRAC (36) and the original PHITS electron track-structure extension (PHITS etsmode) were received. An overview of the participants' codes is given in **Table 3**, together with information about the energy cut-off used in the simulation for electron transport.

The frequency means $\bar{z}_f$ of the revised and new results for the specific energy distribution in the target sphere with a surface source are given in **Table 4**. Results already published in ( 4) concerning the frequency-mean specific energy per decay for the other two configurations are also included in the table. For illustrating the results of this benchmark, a supplementary figure (see Fig. S2) shows the differences in the microdosimetric spectra for the three configurations obtained with Geant4-DNA default option.

**Table 4:** Monte Carlo codes used for the microdosimetric calculations and the frequency-mean specific energy obtained for the surface source configuration. With a view to complementarity, we also include the values that were obtained in the point and volume configurations that have been published previously ( 4). We add here results obtained with PHITS etsmode that have been recently calculated.

| Monte Carlo code/option | Frequency-mean specific energy $\bar{z}_f$ per decay (mGy) | | |
|---|---|---|---|
| | surface source (this work) | point source | volume source |
| **Geant4-DNA Opt0/default (V10.3)** | 2.06 (7.4 eV ; participant 1) ; 2.07 (14 eV ; participant 2) | 4.11; 3.96 | 3.83; 3.69 |
| **Geant4-DNA Opt2 (V10.3)** | 2.02 | 4.07 | 3.80 |
| **Geant4-DNA Opt 7 (V10.3)** | 1.98 | 3.98 | 3.74 |
| **PENELOPE** | 3.55 | 4.08 | 3.91 |
| **MNCP6.1** | 3.08 (20 eV, participant 3) ; 1.62 (14 eV ; participant 2) | 3.99; 3.91 | 3.76; 3.75 |
| **FLUKA** | 3.16 | 4.04 | 3.85 |
| **NASIC** | 2.07 | 4.13 | 3.83 |
| **PHITS** | 2.09 | 4.13 | 3.86 |
| **PHITS etsmode** | 2.01 | 4.01* | 3.72* |
| **PARTRAC** | 2.21 | - | - |
| **Mean value and standard deviation** | 2.3 ± 0.6 | 4.04 ± 0.07 | 3.80 ± 0.07 |

*Not previously published

The mean values of the specific energy in the sphere shown in **Table 4** allow to quickly check the validity of the calculations. Indeed, if all the energy released per decay of the source was imparted inside the sphere, the expected specific energy will be around 5.97 mGy. In the continuous slowing-down approximation (CSDA), only electrons with kinetic energy larger than ~15 keV have a range in liquid water exceeding the radius of the sphere ( 37). Considering their emission probability in the spectrum of the artificial source (see Supplementary Fig. S3(a) and (b)), the value of 4.04 mGy for the point source configuration is in qualitative agreement with the energy found in the contributing part of the spectrum. For the surface configuration, one can expect roughly half of the energy deposited for the point source and a value in between these two extreme cases for the volume distribution as can be seen in Table 4.

To corroborate this qualitative argument, the energy imparted by emitted electrons within the target sphere was estimated based on the electron CSDA ranges (see Supplementary Fig. S3(b)) and the chord length distribution (Supplementary Fig. S3(c)) for the three considered geometries. The majority of the emitted electrons have energies below 10 keV which is the lowest energy given in the ESTAR (electron stopping power and ranges) data base of the National Institute of Standards and Technology (NIST) ( 37). For lower energy electrons, the CSDA ranges obtained by extrapolation of the data at higher energies can be expected to be an upper limit for the actual distance from a given start point that electrons travel before being stopped. This extrapolation was based on a parametric power law $R = \alpha \times (E/E_0)^\beta$ for the range $R$ as function of energy $E$, where $E_0$ was chosen as 10 keV. The best-fit parameters to the data in the ESTAR database for electron energies between 10 keV and 100 keV were $\alpha$ = 2.53 µm and $\beta$ = 1.7586.

Using the mean energy spectrum of emitted electrons shown in Supplementary Fig. S3(a) and aforementioned power law regression, one can determine the mean distributions of the CSDA range and the mean energy imparted within a given distance around the point of electron emission that are shown in Supplementary Fig. S3(b) and S3(d). Convoluting the latter with the chord length distributions for the three geometries gives estimates for the mean specific energy of 3.78 mGy for the volume source, 4.14 mGy for the point source and 2.05 mGy for the surface source. This supports the plausibility of the values shown in the last two columns of Table 4 as well as the results shown for the surface source obtained with track structure codes.

The large spread of values in **Table 4** reveals a mean of 2.3 mGy and a standard deviation of as much as 0.6 mGy. If only track-structure codes are compared, a much better agreement is obtained, with a mean value of 2.04 mGy and a standard deviation of 0.03 mGy. This is also shown in Figure 2, where full microdosimetric spectra obtained with only MC track-structure codes are plotted. In a similar way to what was done in ( 4), a Pearson analysis was performed to quantify the agreement, resulting in most cases in correlation degree > 95%. These large correlation rates must nevertheless be moderated by the fact that several of the codes used share the same physical models (Nasic, geant4-Opt0, Opt2 and Opt0-14 eV) thus increasing their agreement. Nevertheless, the general agreement remains high and, especially, with regard to the results obtained with MC codes that use condensed history approaches, we can conclude that in the case of a source emitting very low energy electrons in this specific configuration (decays taking place on the scoring volume surface) track-structure codes are more suitable.

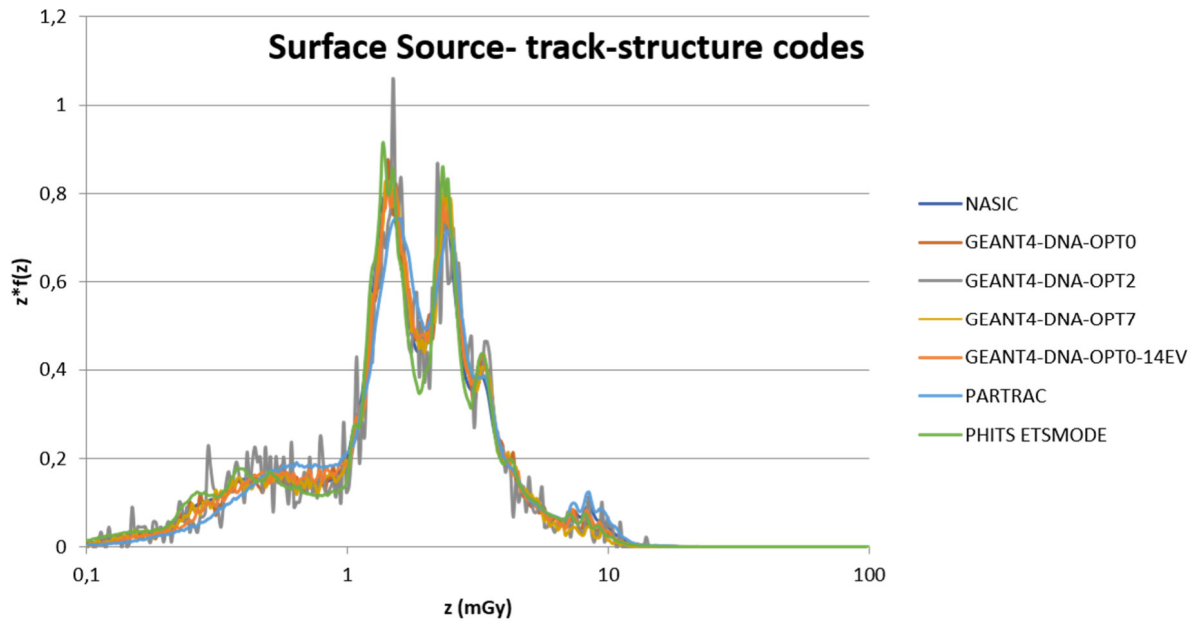

*Figure 2: Comparison of the simulated microdosimetric spectra obtained using only track-structure MC codes for the surface distribution of the $^{125}$I source as defined in ( 23).*

4. **First intercomparison: Nanodosimetric results**

For the nanodosimetric calculations, the participants were asked to score ICSDs in small spherical volumes of 3 nm and 8 nm diameter placed at different distances from the source in a point-source configuration, i. e. at the center of the 5 µm-radius liquid water sphere (Figure 3). As in the microdosimetric simulations, the material used for the entire geometrical set up was liquid water in order to use the same models/cross-section data sets for a given MC code in all volumes.

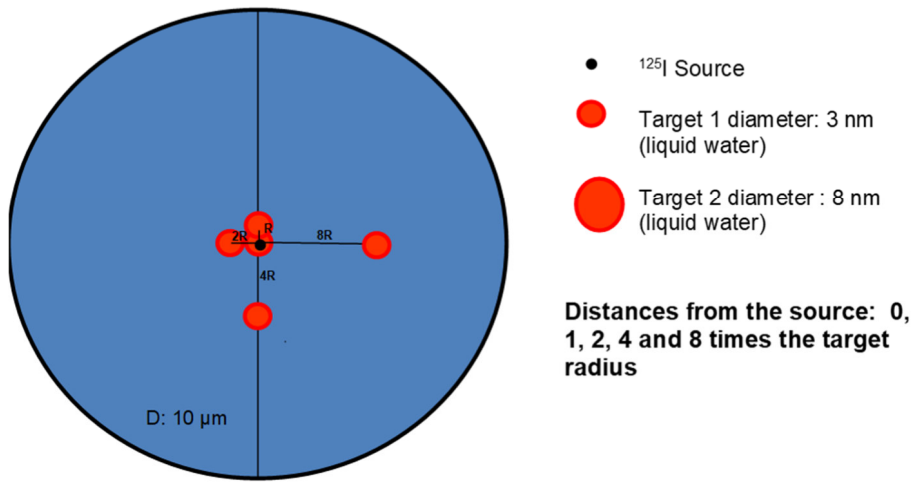

*Figure 3: geometrical setup used for the nanodosimetric intercomparison. The center of target spheres of 3 and 8 nm diameter was placed within liquid water sphere of 10 µm at different distances from the $^{125}$I source (0, 1, 2, 4 and 8 times the target radius) which was placed at the center of the 10 µm diameter sphere. ICSDs were compared for different track-structure codes or options using different cross-sections data sets and models.*

As explained in Section 2, only track-structure codes can be used for the simulation of ICSDs. Therefore, at the time of the first report on the intercomparison exercise, only a few results were reported and a comparison of results obtained with only three different options of Geant4-DNA (version 10.3) was included in (12). Details concerning the different options are as follows: both the default option (option 0) and option 2 use the same physical cross-section models, but their implementation in option 2 is optimized in terms of computational time. For both inelastic cross-sections (ionization and electronic excitation) the Emfiezoglou dielectric model is used (from 9 eV (excitation) or 11 eV (ionization) to 1 MeV) and for elastic scattering, a partial wave model is used from 7.4 eV to 1 MeV. More details about these models can be found in ( 25, 28). Option 7 uses different cross-section data sets than the other two options (i. e. a combination of option 4 models for electrons with kinetic energy < 10 keV and the default option for electron energies above 10 keV). Therefore, in this option, electrons with kinetic energy < 10 keV are transported using the Emfietzoglou-Kyriakou revised dielectric model for electron energies from 8 eV for the electronic excitation process and 10 eV for the ionization process. The Uehara screened Rutherford model is used for calculating the elastic scattering cross-section ( 25, 38). The difference in the ICSDs obtained with these two groups of cross-sections was surprisingly large, especially for volumes containing the source, which received on average a higher number of ionizations. Differences were as much as 24% in the mean and up to 16% in the standard deviations of the ICSDs.

After the first analysis was completed, new participants sent their results for ICSDs obtained from simulations with track-structure codes. This new data confirmed the first insights from the previous analysis: ICSDs calculated with different track-structure codes using different physical models for describing both inelastic and elastic interactions show large variations. Details about the models used for each code are given in their corresponding reference papers: MDM/LQD ( 39), PARTRAC ( 36), Geant4-DNA option 6 ( 25, 40) and PHITS etsmode( 32). These variations are reflected by the scatter of the distribution mean values reported in **Table 5**, where results for a sphere of 3 nm diameter containing the $^{125}$I point source are presented. Similarly, a large scatter was observed for target volumes at other distances from the source and ICSDs obtained for an 8 nm-diameter sphere.

**Table 5.** MC Track-structure codes for the nanodosimetric calculations and the mean ionization cluster size obtained for a 3 nm-diameter target sphere with the electron source at its center. The number of simulated histories varies between different participants/codes but was sufficiently high to obtain ICSDs with negligible statistical uncertainty.

| Track-structure code/option | Mean value of the ICSDs |
| --- | --- |
| **Geant4.10.2 opt2** | 16.40 |
| **Geant4.10.2 opt7** | 14.09 |
| **Geant4.10.4 opt6** | 26.34 |
| **MDM (formerly LQD)** | 11.08 |
| **PHITS etsmode** | 7.52 |
| **PARTRAC (D)*** | 12.69 |
| **PARTRAC (A)*** | 14.36 |
| **PARTRAC (TS)*** | 10.13 |
| **Mean value± sample standard deviation** | 14±6 |

* For PARTRAC: (D) default option, approximation of low-energy electron transport below 30 eV; (A) default option + auto-ionization events occurring with 50-55% probability for 4 (of 5) classes of excited states of water (41) are also taken into account when scoring ionizations; (TS) full event-by-event simulation down to 10 eV.

Taking advantage of the flexibility of Geant4-DNA as it offers different physical models and options as well as being open-source, a preliminary sensitivity study was performed and reported in the publication (12). The total interaction cross-section was changed by ±10% in both options (options 0 and 2), and the effect on the calculated ICSDs was evaluated. Such a large magnitude of variation in the total interaction cross-section had, however, a comparatively small effect on the mean values of the ICSDs. This indicated that larger variations in the cross-sections were needed in order to obtain a noticeable effect. In particular, it was observed that changing only the total interaction cross-section by a factor of two altered the results obtained with the default physical models in Geant4-DNA in such a way that the ICSDs were similar to those obtained with another track-structure code, MDM (formerly LQD) ( 39), which uses rather different physical models for describing both the inelastic and the elastic interactions of electrons (see **Figure 4**). Of course, modifying the total cross-section by such a large factor is not realistic and leads to changes in the stopping power or the values of *S*-factors that would not be in accordance with the existing data that have been used for the validation of the cross-section models used in the codes. Such modifications, however, provide a rough estimate of the extent to which the variation of cross-sections can impact the physical observable quantities (i. e. ICSDs), which is important for assessing the impact of uncertainty on cross-sections.

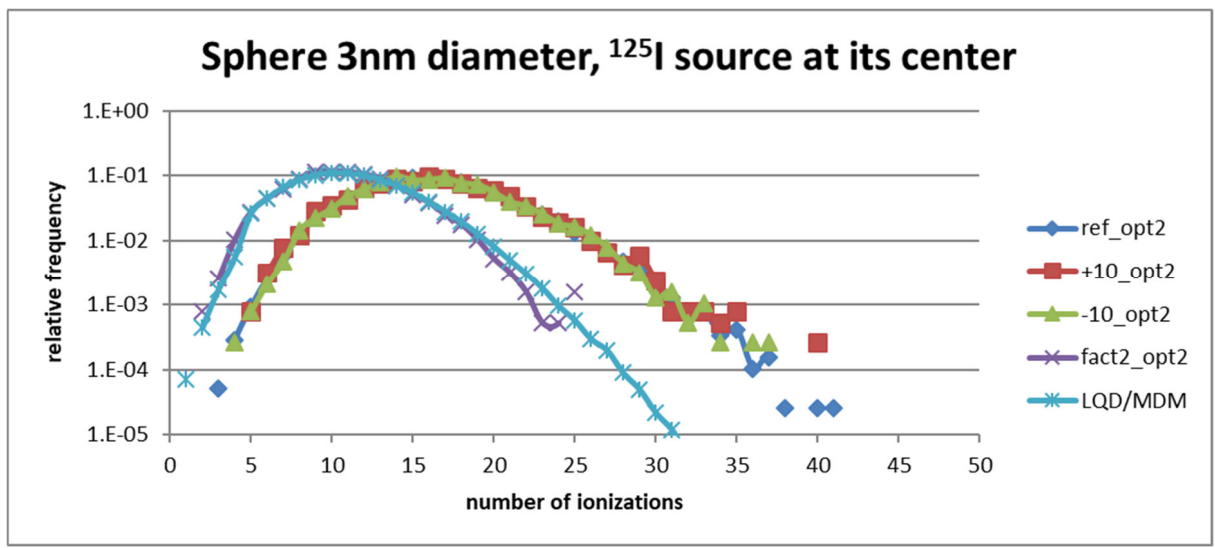

*Figure 4: CSDs obtained for a 3nm-diameter liquid water sphere containing the $^{125}I$ source at its center. Geant4-DNA opt 2 physical models were used with the following configurations: no modification (ref_opt2); ± 10% (±10_opt2) in the total interaction cross-section and decreasing it by a factor of 2 (0R-fact2_opt2). These results are compared to the ICSDs obtained with the MDM (formerly LQD) track-structure code with no modifications Note: Lines connecting the points in the ICSDs serve as a guide for the eyes since only integer numbers are possible.*

### 5. Sensitivity analysis on the impact of inelastic cross-section modifications on ICSD results

In this Section, we present an extension of the sensitivity analysis to further investigate how variations in the electron inelastic cross-sections can affect the results obtained in terms of ICSDs. It is well known that physical models used in track-structure codes for describing these interactions agree quite well among each other and with experimental data for electrons of kinetic energy > 10 keV. At lower energies, deviations arise between different cross-section data, which generally increase with decreasing electron kinetic energy. As mentioned in Section 2, some authors claim (42, 43), that cross-sections down to 100 eV have been derived with theoretical uncertainties between 5% and 10% which is lower than the uncertainty of experimental data used for their validation that range between 20% and 40%. These low uncertainties are of the same order of magnitude as those recommended by the ICRU or NIST databases for stopping powers of electrons with energies from 1-10 keV. Nevertheless, for even lower energies, the uncertainties involved in the calculations significantly increase, and the relative spread of inelastic cross-sections reported in the literature for electrons down to 10 eV exhibits discrepancies of as much as 20–100%, or in some cases even more (44).

It, therefore, seems logical to consider within the sensitivity analysis modifications in the inelastic cross-sections that depend on the electron energy. For this study, we adopted the following expression:

$\sigma^* = \sigma\,[1+f(E)]$ (1)

where, $\sigma$ is the cross-section as implemented in the code, $\sigma^*$ is the modified cross-section and $f(E)$ is an energy-dependent relative change. The following two energy dependencies were heuristically chosen:

$$f_1(E) = -0.1752 \ln (E/\text{eV}) + 1.4477 \tag{2}$$

$$f_2(E) = -0.1473 \ln (E/\text{eV}) + 1.1495 \tag{3}$$

For an electron energy E of 10 eV, these functions result in a relative change of about 100% and 80% for $f_1$ and $f_2$, respectively, where both values decrease with increasing electron energy as shown in Figure 5.

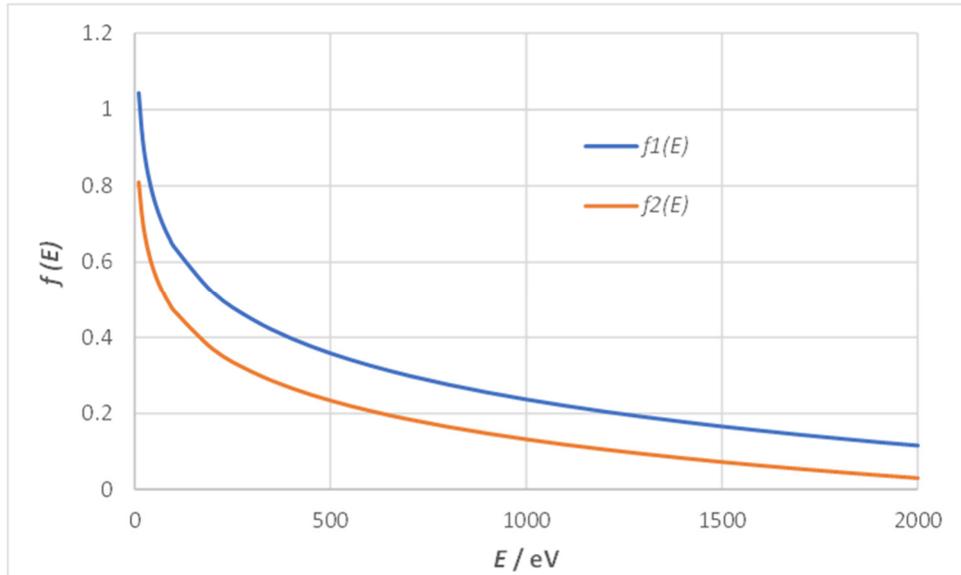

Figure 5 Representation of the functions $f_1(E)$ and $f_2(E)$ used in the modification of the inelastic cross-section in Geant4-DNA option 4 as a function of the electron energy for electrons with < 1 keV. (For details see text).

These functions were used in two separate trials that modified the inelastic scattering cross-sections with option 4 in Geant4-DNA (and applied only for electrons with kinetic energy less than 1 keV). The first trial consisted the use of $f_1$ for $f$ in equation (1). In the second one, $f_2$ and $f_1$ were used as lower and upper bounds, respectively, for a relative change that had a random variation with energy, i.e.

$$f(E) = f_2 + (f_1 - f_2) \times \text{rand}(0,1) \tag{4}$$

where rand(0,1) is a random number between 0 and 1, and $f(E)$ is then used in equation (1). Figure 6 shows the results obtained with Geant4-DNA using the default option (Opt 0), option 4 (Opt 4) and option 4 combined with the modifications based on the two previously mentioned functions $f_1$ and $f_2$ for cross-section modifications when calculating ICSDs in a liquid water sphere. In this case, results are shown for an 8 nm-diameter target using monoenergetic electron sources of 50, 200 and 1 keV initial electron energy placed in its center. Similar observations were made for the smaller target (3 nm diameter) and other intermediate energies.

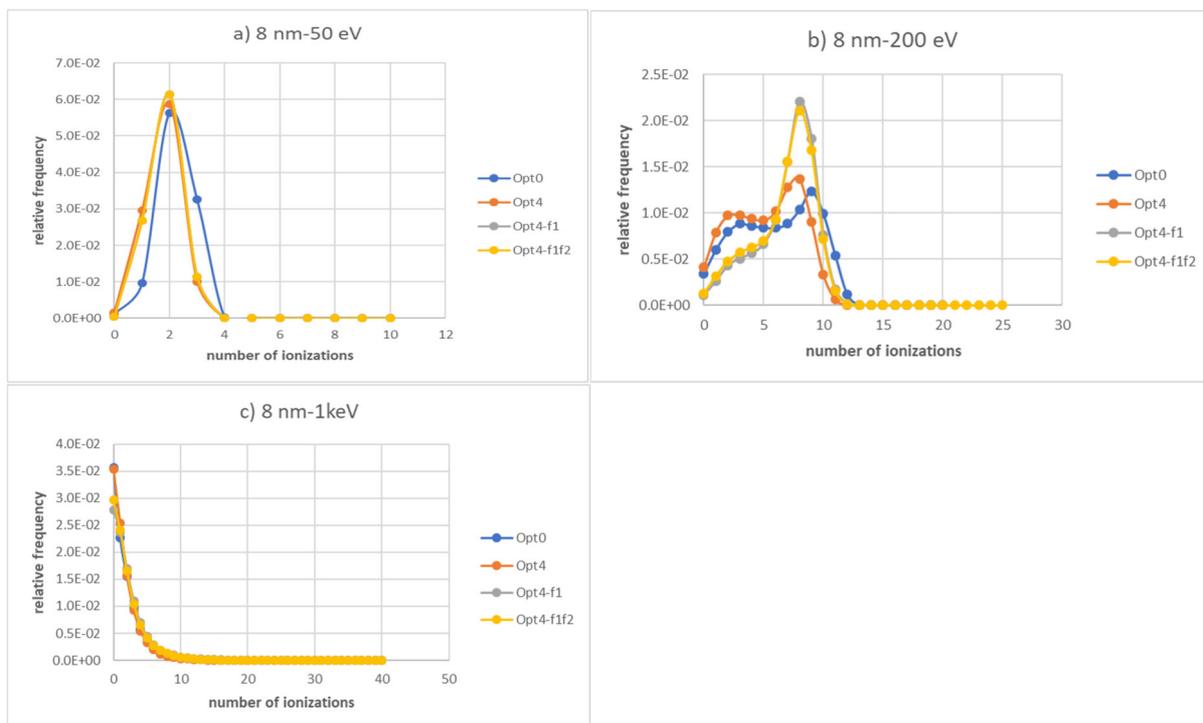

*Figure 6: ICSDs produced by monoenergetic electrons with initial energy of (a) 50 eV, (b) 200 eV and (c) 1 keV in 8 nm-diameter liquid water spheres. Results using Geant4-DNA default option (Opt0) and Geant4-DNA option 4 (Opt4) are compared with those obtained after a modification of the Option 4 inelastic cross-sections with the functions $f_1(E)$ and $f_2(E)$ as described in the text. Note: Lines connecting the points in the ICSDs serve as a guide for the eyes since only integer numbers are possible.*

The electron binding energies of the molecular shells of water used in Geant4-DNA are 10.79; 13.39; 16.05; 32.3 and 539 eV for the default option and 10; 13; 17; 32.2 and 539.7 eV for Opt4. Therefore, for monoenergetic electrons of 50 eV, no more than 4 ionizations can be obtained and for 200 eV no more than around 15. For 1 keV electrons, more ionizations are possible but at this energy the cross sections are lower and most of the interactions of the primary particle will not be in the target volume. On the other hand, secondary electrons also interact in the volume. In fact, the product of IMFP for ionization by the primary electron (see Supplementary Figure S4) multiplied by the sphere radius is about unity resulting in a probability for an ionizing interaction of the primary electron within the 8 nm-diameter sphere of about 60%. Since the ICSD drops much slower than would be expected from these number, secondary electrons produced outside the volume contribute significantly to the scored number of ionizations in the target at higher energies.

From the results shown in **Figure 6**, it can be seen that, when modifying the cross-sections, the use of a constant relative shift $f_1(E)$ or of a random value of the relative shift between $f_2(E)$ and $f_1(E)$ has a small impact on the resulting ICSDs. After modifications in the inelastic cross-sections with option 4, the shape of the results tends to be preserved but the mean value noticeably shifts towards values closer to that of the ICSD obtained using the default option. Most importantly, the standard deviations of the new ICSDs increased compared to the distributions obtained with unmodified cross-sections, which may be related to the artificial nature of such modifications.

It can also be seen that the modified cross sections result in smaller changes of the ICSD for 50 eV initial electron energy compared to changes for 200 eV initial energy. This appears paradoxical since the cross sections for electron-impact ionization at the two energies are almost the same (cf. Supplementary Figure S4) and a larger modification is applied to the 50 eV cross section. In addition, also the proportion of ionizations producing secondary electrons with energies high enough to be also ionizing is similar for the two energies (as can be seen from the cyan curve compared to the solid black one in Supplementary Figure S4). However, it must be noted that the cross section for elastic scattering (green curves in Supplementary Figure S4) is by a factor of 5 higher than the ionization cross section at 50 eV, whereas it is only about 70% of the ionization cross section at 200 eV (for the data from Geant4 opt4 and the Champion elastic cross sections that are shown in Supplementary Figure S4). Therefore, increasing the ionization cross section has a minor effect on the frequency of ionizations at 50 eV, since the transport of the electrons is governed by the random-walk like trajectory imposed by elastic scattering. For 200 eV (and also 1 keV), on the other hand, the frequency of ionizations is expected to increase as ionization is the dominant interaction process.

Similar observations as in **Figure 6** can also be made in **Figure 7**, where the ICSD of the previously defined electron "$^{125}$I source" is calculated in the same manner as described above for the monoenergetic electrons. In this case, Geant4-DNA Option 7 is used as option 4 inelastic cross-sections are only valid for electrons with energies up to 10 keV. In option 7, option 4 is thus used for electrons up to 10 keV, while electrons with higher energies are transported using the default-option cross-sections. Only the inelastic cross-sections of Option 4 for electrons with energy < 1 keV were modified with the *$f_1(E)$* and *$f_2(E)$* factors.

The mean number of ionizations of the distributions shown in **Figure 7** of between 30 and 40 is in the range that is expected from the estimated energy imparted in the continuous slowing down approximation (that is of limited validity for the low energy range below 1 keV). From the inset of Supplementary Fig. S3, the mean energy imparted per decay within a 4 nm sphere is about 1.35 keV. Since, the lowest energy in the spectrum (corresponding to the CK-OOX transitions in $^{125}$I) is 6 eV, these electrons are not ionizing, so that one has to subtract the 0.55 keV deposited within the range of these electrons from the total energy imparted to obtain about 0.8 keV imparted by electrons capable of ionizing interactions. With a mean energy per ionization of about 20 eV, this suggests a mean number of ionizations of about 40.

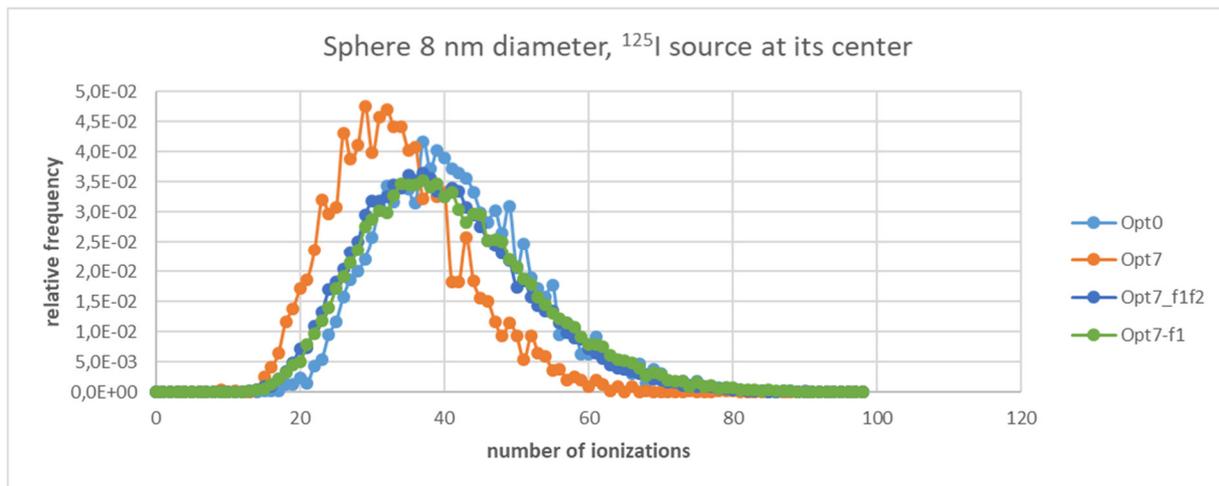

Figure 7: ICSD results in an 8 nm-diameter liquid water sphere calculated with Geant4-DNA default option (Opt0), unmodified option 7 (Opt7) and option 7 modified with the functions f1 and f2 as described in the text. Note: Lines connecting the points in the ICSDs serve as a guide for the eyes since only integer numbers are possible.

### 6. Discussion: defining an approach to determine the impact of cross-section data sets on the dispersion of ICSDs results.

The analysis of results in this work represents a preliminary study, which was necessary for establishing a range for the variation of the inelastic cross-sections to be further used by participants working with different track-structure codes, with the aim of investigating the impact of such modified cross-sections when calculating nanodosimetric distributions.

In general, when designing a code comparison, care should be taken to minimize the possibility of errors when participants are asked to use the same input and to make the final analysis as simple as possible for the organizers and less error-prone. With this in mind, the following conclusions could be drawn from this work, for the next step of the exercise:

The standard approach would consist in asking participants to perform calculations (in this case of ICSDs) varying the cross-sections implemented in their codes and then compare their results. For this purpose, participants could repeat the ICSDs calculations obtained in the first part of the intercomparison (4), but with modified inelastic cross-sections following a variation procedure established by the organizers. This would constitute a sensitivity analysis for each code. The idea should always be that the participants perform as few and as simple changes to the code as possible. This is to limit errors and to keep the exercise open to as many participants as possible, including those who are not among the main developers of a code but relatively advanced users.

With this approach in mind, functions for variation of cross-sections depending on electron energy could be proposed, such as the ones tested in this work ($f_1(E)$ and $f_2(E)$). However, the logarithm dependence might not be well suited as the logarithm is not limited, and thus, the relative change of the cross-section changes sign (for the function $f_1$, this will occur at an energy of 3.8 keV). Therefore, if a smooth and monotonously decreasing energy dependence is to be described by a simple function, alternatives like an inverse power law, $f(E) \propto 1/E$, are preferable.

More generally, $f(E)$ should be independent on the cross-section data set, so that it can be applied to any code. Ideally, $f(E)$ should be chosen in accordance with recent estimations of the uncertainties by different theoreticians for the cross-sections used in the codes (either with cross-section model functions or data in lookup tables).

Furthermore, it would also be beneficial to study incremental changes in the modifications of the cross-sections. For example, participants could be asked to modify their inelastic cross-sections according to

$$\sigma(E)_{test\ i}^{+/-} = \sigma(E) \times [1 + A \times f(E)]^{i/10} \tag{5}$$

where i$= -10, -9, \ldots, 1, \ldots, 10$, thus both decreasing and increasing the original cross-section value and where *A* is the maximum relative offset (at a reference energy $E_0$). This allows exceeding 100% variation in the positive direction whilst keeping positive cross-section values.

The main problem faced when trying to determine the uncertainty in the ICSDs attributed to different cross-sections is, of course, that all these cross-section data sets are traceable to experimental data and have been validated within the limits explained in Section 2. Thus, there is no prior criterion for the "best" or the "true value" of a given cross-section data set (for both inelastic and elastic interactions), and there is also not an *a priori* best ICSD to use as benchmark for the other results. Therefore, following the approach of the GUM, a "consensus" ICSD is needed to evaluate the deviation of the results for the different codes. After introducing step-changes in cross-sections according to the proposed functions, quantitative measures can then be used for the differences in pairwise comparisons between participants' results. The consensus ICSDs and underlying cross-section dataset could be the ones minimizing such differences.

Such a procedure has some complexity associated with it, as it originates from the attempt to infer the consensus ICSD with respect to a given 'parameter' (i.e. the cross-sections used) by varying this parameter using discrete steps. This procedure would require the participants to perform a large number of simulations (thus increasing the risk of possible errors), which may also introduce a bias in the consensus ICSD depending on the number of discrete steps used for the study.

An alternative approach would be to obtain the consensus ICSD directly by using a "consensus" cross-section data set. To this end, participants would need to provide their inelastic and elastic cross-sections for a pooled analysis to produce a table of mean values of the cross-sections at each considered value of independent variables (projectile energy and, if applicable, scattering angle, energy loss, secondary particle energy and emission angle).

This procedure has also the advantage that, individual cross-sections of the codes can be first compared (including total and differential cross-sections for inelastic and elastic scattering) using the same format. This direct comparison of the cross-sections will be used to identify groups related to the physical models used. Using the framework of the GUM, these groups will have to be considered as data with correlated uncertainties. By considering these correlations, a better estimate of the "true" value cross-section than an arithmetic average of the data can be obtained. Furthermore, this direct comparison will enable an empirical test of cross-sections based on extrapolation only compare with those from a plausible theoretical approach and experimental validation, which would, a priori be believed to be more credible. However, this needs to be assessed when all the cross-sections implemented in the participants' track-structure codes are compared. For different monoenergetic electrons, participants would then calculate the corresponding ICSD using these consensus cross-section tables. Finally, the uncertainty associated with a particular cross-section data set can be determined from the deviation between the ICSD obtained with this particular data set and the consensus ICSD.

## 7. Conclusion and perspectives:

This paper completes the results presented in (4) concerning the intercomparison of different MC codes for the determination of microdosimetric spectra for different source configurations. In the case of very low-energy electrons started at the surface of the target volume, the use of condensed history approaches together with an energy cut larger than 50 eV have been shown to introduce errors (artefacts) in the calculation of microdosimetric spectra that are not present when track-structure codes are used.

MC track-structure codes are the only ones that can be used to calculate nanodosimetric quantities, namely the ICSDs. In the codes, different physical models are implemented for electron interaction cross-sections at very low energy for both inelastic and elastic interactions. This seems to be the origin of significant differences in the nanodosimetric results even though all these cross-section data have been validated by benchmarking calculations of measurable quantities on larger spatial scales, such as the stopping power, particle ranges or $S$-factors. These discrepancies at the nanometric scale lead to the question of what uncertainty is to be attributed to the simulated results. More precisely and quantitatively, the question is the magnitude of the uncertainty contribution arising from the cross-sections used by the codes, which should be disentangled from the contribution of other code features such as the tracking procedure or the interpolation methods used for the calculations of the path length, etc. In order to answer these questions, this work studied the variability of ICSDs simulated with different cross-sections data sets. Two different procedures have been tested that modified the inelastic cross-section dataset using the Geant4-DNA option 4 and an energy-dependent function that took into account the fact that uncertainties are larger at lower electron energies. From the above discussion of the limitations of such an approach, it was concluded that the contribution of cross-sections to the uncertainty in nanodosimetric results is difficult to achieve from a comparison of ICSDs obtained with different codes due to the difficulty of defining a consensus result among the codes. The proposal of a pooled analysis of different cross-section data to establish a mean/consensus data set is a promising alternative for estimating uncertainties associated with simulated micro- and nanodosimetric quantities. With this goal in mind, the second part of this activity is currently underway, where users of the MC track-structure codes are welcome to participate.

**Acknowledgements:**

Werner Friedland is acknowledged for providing the simulation results with the PARTRAC code.

**Supplementary Material to "Intercomparison of micro- and nanodosimetry Monte Carlo simulations: an approach to assess the influence of different cross-sections for low-energy electrons on the dispersion of results" by Carmen Villagrasa, Hans Rabus, et al. in Radiation Measurements**

**Supplementary Figure 1**

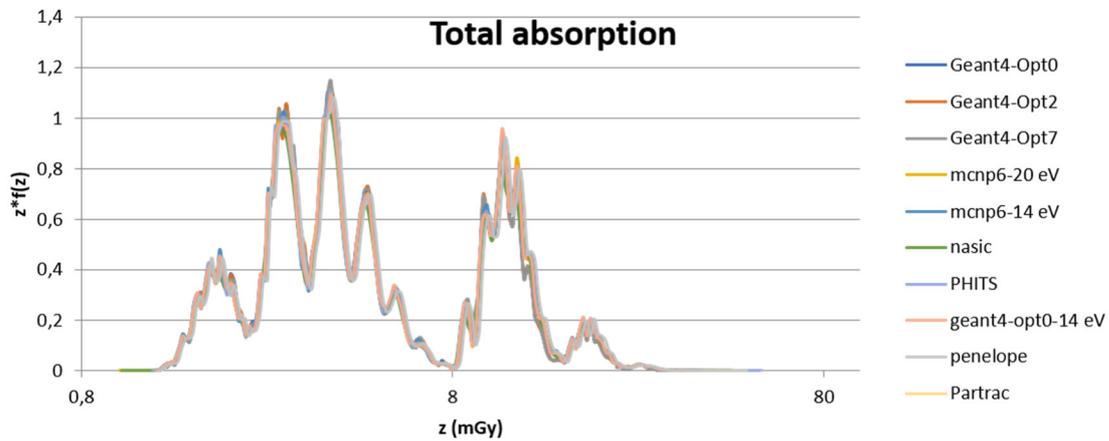

*Supplementary Figure S1: Comparison of the spectra of energy released per decay reported by the participants (normalized to the mass of the 5 µm-radius water sphere).*

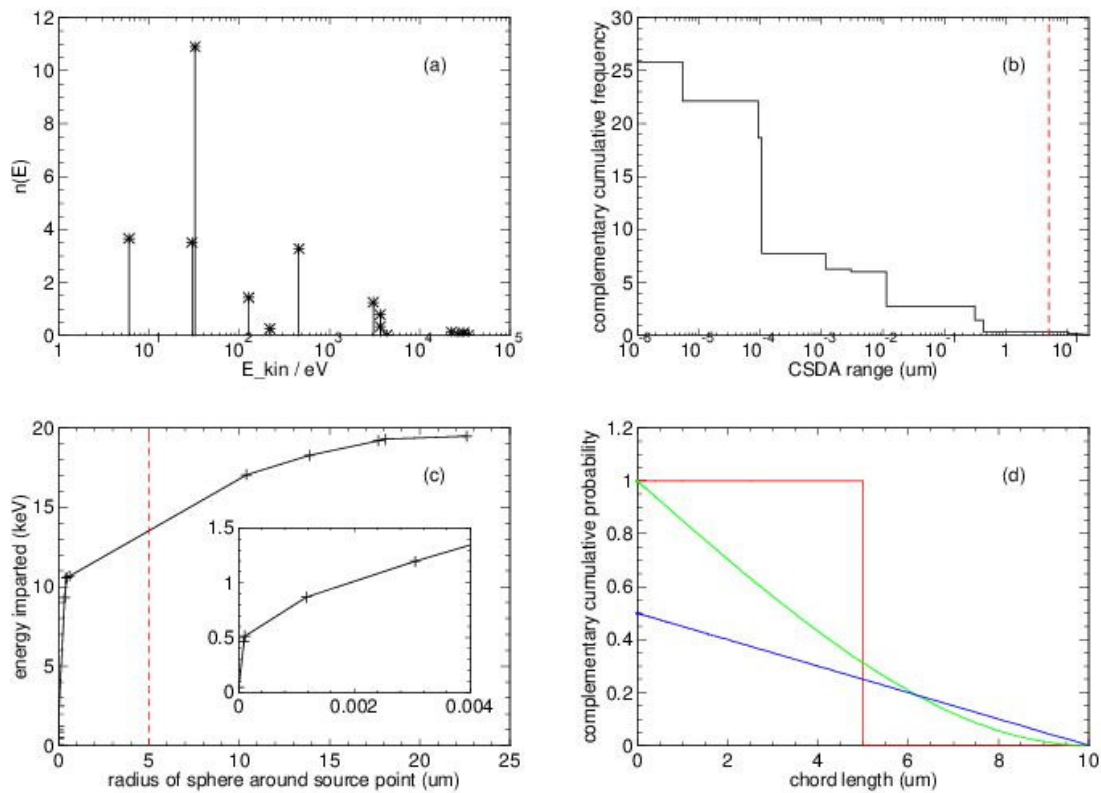

*Supplementary Figure S2: (a) Average number of electrons emitted in a decay of the source. (b) Complementary cumulative distribution of the average number of electrons emitted from the source used in the exercise that have a CSDA range exceeding the value given on the x-axis.(c) Average energy imparted in a sphere around a source point under CSDA conditions (all emitted electrons travel along straight lines with constant energy loss per unit length). The inset shows a close-up on radii up to 4 nm, i.e. radius of the larger of the target spheres used in the nanodosimetric part of the exercise. (d) Complementary cumulative chord length distributions for the point source (black), the volume source (green) and the surface source (blue). The plotted values are the probabilities of chord lengths exceeding the value on the x-axis. The vertical dashed lines in (b) and (c) indicate the sphere radius. The respective point on the solid line in (b) is, therefore, the mean energy imparted in the sphere for the case of the point source (that corresponds to a specific energy per decay of 4.14 mGy).*

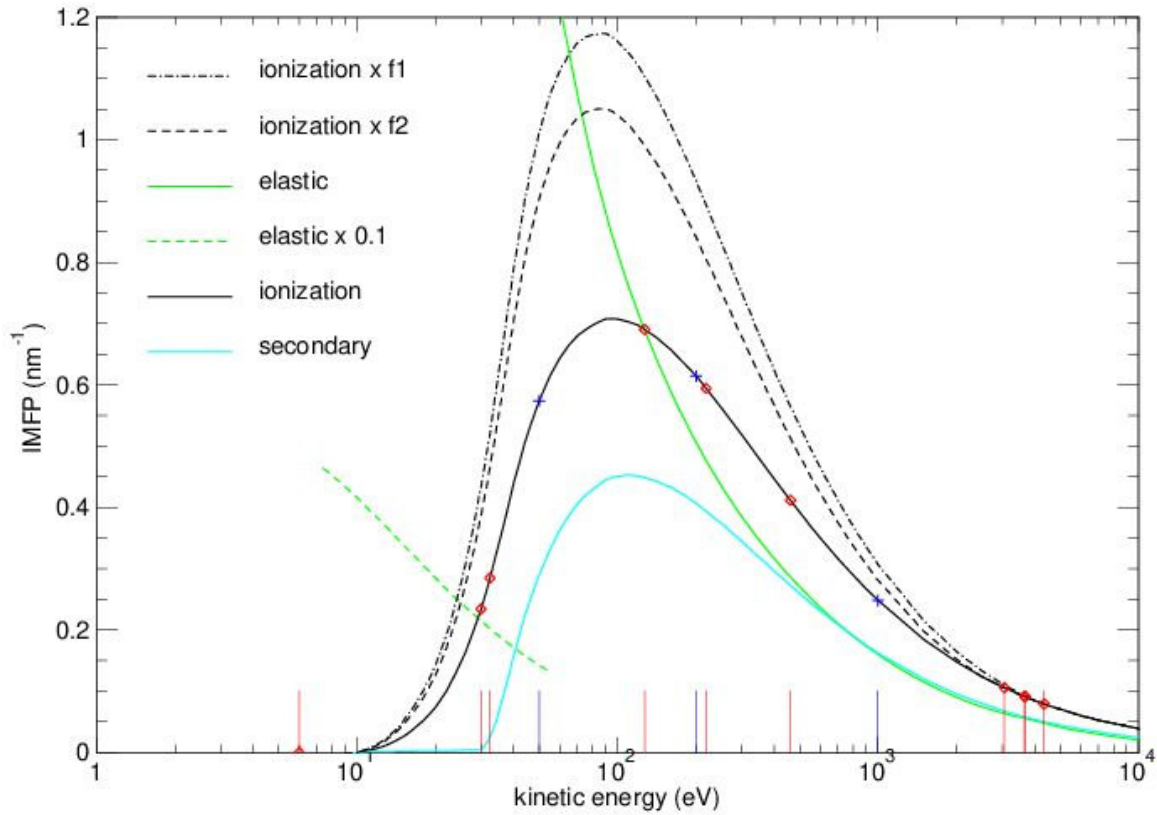

*Supplementary Figure S3: Inverse mean free path (IMFP), i.e. the cross sections multiplied by the number density of scatterers, for electron interactions as a function of the kinetic energy. The data apply to Geant4-DNA option 4 with Champion's cross sections for elastic scattering. The dot-dashed and dashed black lines represent the upper and lower bounds of the modified ionization cross-sections. The vertical blue lines indicate the energies used for the data shown in Fig. 6 of the paper, the vertical red lines indicate the energies of the artificial electron source used in the exercise. The dashed green line shows the elastic scattering cross-section at energies below about 55 eV that has been multiplied by a factor 0.1 such as to fit the vertical scale.*